\documentclass[journal]{IEEEtran}

\usepackage{array}
\usepackage{booktabs}
\usepackage{graphicx}
\usepackage{threeparttable}
\usepackage{multirow}
\usepackage{subcaption}
\usepackage{cite}
\usepackage{stfloats}
\usepackage{amsmath}
\usepackage{amssymb}
\usepackage{url}
\usepackage[labelfont=bf, labelsep=period, font=small]{caption}
\ifCLASSINFOpdf
\else
\fi

\begin{document}
\title{seg2med: a bridge from artificial anatomy to multimodal medical images}
\author{Zeyu~Yang,
        Zhilin~Chen,
        Yipeng~Sun,
        Anika~Strittmatter,
        Anish~Raj,
        Ahmad~Allababidi,
        Johann S.~Rink,
        and~Frank G.~Zöllner,~\IEEEmembership{Senior~Member,~IEEE}
\thanks{Z. Yang, Z. Chen, A. Strittmatter, A. Raj, and F. G. Zöllner are with the Computer Assisted Clinical Medicine, Medical Faculty Mannheim, Heidelberg University, Mannheim, Germany, e-mail: zeyu.yang@medma.uni-heidelberg.de, anika.strittmatter@medma.uni-heidelberg.de, anish.raj@medma.uni-heidelberg.de, frank.zoellner@medma.uni-heidelberg.de).}
\thanks{Y. Sun is with the Pattern Recognition Lab, Friedrich-Alexander-University Erlangen-Nuremberg, Erlangen, Germany.}
\thanks{A. Allababidi and J. Rink are with the Department of Radiology and Nuclear Medicine, University Medical Center Mannheim, Mannheim, Germany, e-mail: johann.rink@umm.de}
\thanks{Z. Yang, Z. Chen, A. Strittmatter, A. Raj, and F. G. Zöllner are with the Mannheim Institute for Intelligent Systems in Medicine, Medical Faculty Mannheim, Heidelberg University, Mannheim, Germany}
\thanks{Z. Chen is with the Optical Bioimaging Laboratory, Department of Biomedical Engineering, College of Design and Engineering, National University of Singapore, Singapore, e-mail: zhilin.chen@u.nus.edu}%
}

\maketitle

\begin{abstract}
We present seg2med, a modular framework for anatomy-driven multimodal medical image synthesis. The system integrates three components to enable high-fidelity, cross-modality generation of CT and MR images based on structured anatomical priors.

First, anatomical maps are independently derived from three sources: real patient data, XCAT digital phantoms, and anatomies-synthetic subjects created by combining organs from multiple patients. 
Second, we introduce PhysioSynth, a modality-specific simulator that converts anatomical masks into imaging-like prior volumes using tissue-dependent parameters (e.g., HU, T1, T2, $\rho$) and modality-specific signal models. It supports simulation of CT and multiple MR sequences, including GRE, SPACE, and VIBE.
Third, the synthesized anatomical priors are used to train 2-channel conditional denoising diffusion probabilistic models (DDPMs), which take the anatomical prior as a structural condition alongside the noisy image, enabling it to generate high-quality, structurally aligned images within its modality.

The framework achieves a Structural Similarity Index Measure (SSIM) of $0.94\pm0.02$ for CT and $0.89\pm0.04$ for MR images compared to real patient data, and $0.78\pm0.04$ FSIM for simulated CT from XCAT. The generative quality is further supported by a Fréchet Inception Distance (FID) of 3.62 for CT synthesis. In modality conversion tasks, seg2med attains SSIM scores of $0.91\pm0.03$ (MR→CT) and $0.77\pm0.04$ (CT→MR).

In anatomical fidelity evaluation, synthetic CT images achieve a mean Dice coefficient exceeding 0.90 for 11 key abdominal organs, and over 0.80 for 34 of 59 total organs. These results underscore seg2med’s utility in cross-modality image synthesis, dataset augmentation, and anatomy-aware AI development in medical imaging.

A web-based API demo is available at: \url{https://huggingface.co/spaces/Zeyu0601/frankenstein}

\end{abstract}


\section{Introduction}
\IEEEpeerreviewmaketitle
\IEEEPARstart{C}{omputed}  tomography (CT) and magnetic resonance imaging (MRI) are two important modalities in medical imaging. MRI is particularly effective for visualizing soft tissue structures with high contrast, while CT offers complementary insights with precise visualization of bone anatomy and essential electron density information for accurate dosimetric calculations in radiotherapy.

Due to their unique strengths, combining CT and MRI in multimodal imaging has become routine in clinical practice \cite{3476-diffusionconversion}. This integration enhances diagnostic precision in abdominal and pelvic imaging, improving assessment accuracy across a range of conditions. For example, in total hip arthroplasty (THA) patients, CT accurately measures lesion size in osteolysis assessments, while MRI is better suited for detecting small lesions in challenging locations. Together, they provide a complete assessment, balancing MRI’s sensitivity with CT’s volumetric precision \cite{Osteolysis}. Similarly, in liver cancer imaging, CT serves as an initial tool to survey metastases, while MRI with liver-specific contrast agents offers additional diagnostic clarity for small or indeterminate lesions. This combination supports comprehensive characterization and staging, especially for patients with chronic liver disease or those preparing for surgery, as CT provides anatomical details and MRI enhances soft tissue contrast \cite{LiverCancer}.

Despite their complementary roles in medical imaging, the lack of paired CT and MRI data remains a significant challenge, as these scans are typically performed independently, leading to increased costs, nonrigid misalignment issues, and limited availability of high-quality, matched datasets for advancing medical imaging research. To address these challenges, medical image synthesis has emerged as a promising solution \cite{gomezAPISPairedCTMRI2024, dekeyzerDistinctionContrastStaining2017}.
This technique establishes a mapping function to transform a known source image into a desired target image, effectively enabling cross-modality image synthesis and addressing data scarcity by generating realistic synthetic images \cite{3476-attentionawareMRtoCT}. Recent advances in generative neural networks have led to algorithms that produce realistic high-quality medical images, supporting applications in noise reduction, artifact removal, segmentation, and registration \cite{3476-anikaregistration,3476-rajsegmentation}. 
One common synthesis task is image conversion, typically translating between various medical imaging modalities.
Kalantar et al. developed a framework with UNet, UNet++, and CycleGAN to generate pelvic T1-weighted MRI from CT images. Their CycleGAN model produces high-quality MRI without requiring spatial alignment, improving segmentation, and MRI-only planning for pelvic radiotherapy \cite{kalantarCTBasedPelvicT1Weighted2021}. Similarly, Simkó et al. developed a robust synthetic CT (sCT) generation framework using quantitative maps (proton density, T1, and T2) to enhance generalization across MRI contrasts and scanners. This approach outperformed contrast-specific models and demonstrated effectiveness in radiotherapy planning with varied MRI contrasts \cite{simkoMRContrastIndependent2024}. Baldini et al. compared the pix2pix \cite{3476-pix2pix} and BrainClustering for synthesizing T2-weighted images from T1-weighted inputs, finding that the pix2pix with ResNet achieved the highest accuracy. These synthetic images strengthen segmentation models, particularly in scenarios where specific MRI sequences are missing \cite{baldiniMRIScanSynthesis2024}.

However, these approaches rely on paired multimodal data, which is difficult to obtain, motivating alternative approaches. For instance, Bauer et al. employed a CycleGAN model to generate multimodal synthetic CT and MR images from XCAT phantoms \cite{segars4DXCATPhantom2010}, supporting multimodal registration algorithms \cite{3476-dominikcyclegan}. While GANs, like those used in these studies, generate high-resolution images, they often struggle with training stability and consistency \cite{3476-pix2pix, kalantarCTBasedPelvicT1Weighted2021, 3476-dominikcyclegan, baldiniMRIScanSynthesis2024}. To address these limitations, Denoising Diffusion Probabilistic Models (DDPMs) have emerged, offering stable training and enhanced anatomical detail preservation by progressively denoising images.

Building on these strengths, Han et al. introduced MedGen3D, a DDPM-based framework that generates paired 3D images and segmentation masks for thoracic CT and brain MRI, achieving spatial consistency and anatomical accuracy within single modalities \cite{hanMedGen3DDeepGenerative2023}. Similarly, Dorjsembe et al. presented Med-DDPM, a diffusion model generating high-resolution brain MRIs from segmentation masks, with applications in data anonymization and brain tumor segmentation \cite{dorjsembeConditionalDiffusionModels2024}.

These studies highlight the importance of effective guidance in generative training but reveal key limitations. Most focus on single-modality synthesis, limiting their use in multimodal contexts essential for comprehensive diagnostics. Additionally, reliance on existing datasets can constrain image quality if training data lacks diversity, and current methods still struggle with aligning images across modalities, especially for non-rigid anatomy.

Recently, TotalSegmentator has emerged as a powerful tool for segmenting CT and MRI images,capable of identifying various anatomical structures, including organs, bones, muscles, and vessels. It achieves high accuracy, with a Dice similarity coefficient of 0.943 for CT and 0.824 for MRI, and it adapts to diverse clinical settings while requiring minimal computational resources \cite{wasserthalTotalSegmentatorRobustSegmentation2023, wasserthalTotalSegmentatorMRIDataset2024}. Its applications include organ volumetry, disease characterization, radiation therapy planning, and large-scale radiological studies, making it an ideal choice for advancing medical image segmentation workflows. 
Notably, TotalSegmentator can address the data diversity issue in medical image synthesis by generating unlimited, high-quality anatomical masks from various CT and MR scans. These masks can serve as inputs for synthetic image generation, mitigating the reliance on scarce paired datasets and enabling more robust and generalizable synthesis models. Thus, this study selected TotalSegmentator as the primary segmentation tool for processing real CT and MR images used in this research.

In this study, our main contributions are twofold.

(1) We develop an enhanced segmentation preprocessing pipeline that fuses full-organ masks from TotalSegmentator \cite{wasserthalTotalSegmentatorRobustSegmentation2023} with body contours extracted via OpenCV, producing anatomically faithful segmentation priors that accurately reflect patient-specific anatomy.

(2) We introduce PhysioSynth, a physics-based prior simulation module that converts those segmentation priors into modality-specific physical priors by assigning tissue-level parameters (e.g., HU, T1, T2, $\rho$) and applying MR signal equations. This step goes beyond conventional label-mask inputs, offering fine-grained, interpretable control over modality appearance.

Together, these components enable seg2med to deliver high-fidelity, anatomically aligned cross-modality synthesis with improved flexibility, interpretability, and extensibility.

\section{Methods and Materials}
\label{3476-sec:methods}

\subsection{Datasets}
\label{3476-ssec:data}
In this study, several datasets from different studies were used to add the generalizability of this framework. In summary, 323 CT image volumes were used for training and 59 CT image volumes for testing, 251 MR image volumes were used for training and 20 MR image volumes were used for testing.  56 XCAT phantoms were used for testing only. An overview of the implemented datasets is presented in Table \ref{tabledatasets}.

\subsubsection{SynthRAD2023}
The first dataset is the pelvis dataset of task 1 from SynthRAD2023 challenge with a total of 180 patients with paired MR and CT images in the training dataset, and a total of 30 patients with MR images in the validation dataset \cite{3476-synthrad}. For the MR images, 120 patients in the training set were acquired with a T1-weighted gradient echo (GRE) sequence, and the remaining 60 patients with a T2-weighted SPACE sequence. In the test set, 20 patients were scanned with the T1 GRE sequence and 10 with the T2 SPACE sequence. For further details, please refer to \cite{3476-synthrad}.

For CT images of 180 patients, the data aRecently, TotalSegmentator has emerged as a powerful tool for segmenting CT and MRI images,capable of identifying various anatomical structures, including organs, bones, muscles, and vessels. It achieves high accuracy, with a Dice similarity coefficient of 0.943 for CT and 0.824 for MRI, and it adapts to diverse clinical settings while requiring minimal computational resources \cite{wasserthalTotalSegmentatorRobustSegmentation2023, wasserthalTotalSegmentatorMRIDataset2024}. Its applications include organ volumetry, disease characterization, radiation therapy planning, and large-scale radiological studies, making it an ideal choice for advancing medical image segmentation workflows. 
Notably, TotalSegmentator can address the data diversity issue in medical image synthesis by generating unlimited, high-quality anatomical masks from various CT and MR scans. These masks can serve as inputs for synthetic image generation, mitigating the reliance on scarce paired datasets and enabling more robust and generalizable synthesis models. Thus, this study selected TotalSegmentator as the primary segmentation tool for processing real CT and MR images used in this research.

In this study, our main contributions are twofold.

(1) We develop an enhanced segmentation preprocessing pipeline that fuses full-organ masks from TotalSegmentator \cite{wasserthalTotalSegmentatorRobustSegmentation2023} with body contours extracted via OpenCV, producing anatomically faithful segmentation priors that accurately reflect patient-specific anatomy.

(2) We introduce PhysioSynth, a physics-based prior simulation module that converts those segmentation priors into modality-specific physical priors by assigning tissue-level parameters (e.g., HU, T1, T2, $\rho$) and applying MR signal equations. This step goes beyond conventional label-mask inputs, offering fine-grained, interpretable control over modality appearance.

Together, these components enable seg2med to deliver high-fidelity, anatomically aligned cross-modality synthesis with improved flexibility, interpretability, and extensibility.

\section{Methods and Materials}
\label{3476-sec:methods}

\subsection{Datasets}
\label{3476-ssec:data}
In this study, several datasets from different studies were used to add the generalizability of this framework. In summary, 323 CT image volumes were used for training and 59 CT image volumes for testing, 251 MR image volumes were used for training and 20 MR image volumes were used for testing.  56 XCAT phantoms were used for testing only. An overview of the implemented datasets is presented in Table \ref{tabledatasets}.

\subsubsection{SynthRAD2023}
The first dataset is the pelvis dataset of task 1 from SynthRAD2023 challenge with a total of 180 patients with paired MR and CT images in the training dataset, and a total of 30 patients with MR images in the validation dataset \cite{3476-synthrad}. For the MR images, 120 patients in the training set were acquired with a spoiled T1-weighted gradient echo sequence, and the remaining 60 patients with a T2-weighted SPACE (Sampling Perfection with Application optimized Contrast using different flip angle Evolution) sequence \cite{3476-synthrad}. In the test set, 20 patients were scanned with the T1 GRE sequence and 10 with the T2 SPACE sequence. For further details, please refer togre divided into 170 training samples and 10 test samples. For MR images of 180 patients, this splitting continues to be used: the images of the same 170 patients are used for training and the rest 10 patients with the 30 patients in the SynthRAD-validation dataset, totally 40 patients, are used for testing.

\subsubsection{Internal Abdominal CT}
The second dataset is an internal abdomen-CT dataset (named Internal Abdominal CT) with 163 patients, including 78 patients presenting with aortic dissection(AD) with abdominal extent between 2010/01/01 and 2021/03/01 and 85 patients of AD-negative cases. 
For details on the data set please refer to \cite{Anish_dataset}. This dataset was split into a training set of 153 patients and a test set of 10 patients.

\subsubsection{M2OLIE Abdominal CT/MR}
The third dataset consists of abdominal CT and T1-weighted MR images acquired from 39 patients as part of the M2OLIE project. It includes 48 CT volumes and 91 T1-weighted MR volumes using Volumetric Interpolated Breath-hold Examination (VIBE) sequences in transversal orientation. The MR data comprises 41 VIBE in-phase, 41 VIBE opposed-phase, 4 VIBE Dixon TRA in-phase, and 5 VIBE Dixon TRA opposed-phase volumes. All images were spatially registered using our previously developed convolutional neural network \cite{Anika_dataset}.
Among the MR volumes, 37 in-phase and 37 opposed-phase images were used for training, and 4 in-phase and 4 opposed-phase for testing. Additionally, 3 Dixon in-phase and 4 Dixon opposed-phase volumes were used for training, while 1 Dixon in-phase and 1 Dixon opposed-phase volumes were used for testing. All CT volumes were used exclusively for testing. This dataset is referred to as M2OLIE Abdominal CT/MRI throughout this study.

\subsubsection{XCAT}
The fourth dataset was from the Extended Cardiac-Torso (XCAT) phantom data \cite{segars4DXCATPhantom2010}. The Library of XCAT anatomy files provides an additional 56 adult models (33 males and 23 females) of varying heights (from 153 cm to 186.4 cm) and weights (from 52 kg to 120 kg) at different ages (from 18 to 78). 
For these models, we generate one abdomen CT-XCAT volume and one abdomen MR-XCAT volume for each XCAT model. For CT-XCAT volume, the simulated CT value preset in the XCAT2 software is used \cite{segars4DXCATPhantom2010}. For MR-XCAT volume, the signal equation of the VIBE sequence is used to recalculate the MR signal value so that each slice in volumes has an image intensity value similar to that of the MR image measured by the real VIBE sequence. Since the XCAT-generated masks do not have corresponding Ground Truth images, the XCAT dataset was only used for testing our approach.

\begin{table*}[ht]
\centering
\tabcolsep=0.5cm
\caption{Overview of Implemented Datasets}
\begin{tabular}{llcccccc}
\toprule
\multirow{2}{*}{Dataset} & \multirow{2}{*}{Anatomy} & \multicolumn{3}{c}{CT} & \multicolumn{3}{c}{MR} \\ \cline{3-8}
                 &                  & Total & Train & Test & Total & Train & Test \\ \midrule
SynthRAD CT/MR   & Pelvis           & 180            & 170            & 10            & 210            & 170            & 40            \\ 

Internal Abdomen CT & Abdomen       & 163            & 153            & 10            & 0              & 0              & 0             \\ 
M2OLIE Abdomen CT/MR & Abdomen      & 48             & 0              & 48            & 91             & 81             & 10            \\ 
XCAT             & Abdomen          & 56             & 0              & 56            & 56             & 0              & 56            \\ \bottomrule
\end{tabular}
\label{tabledatasets}
\end{table*}

\subsection{Multimodal Data Processing}

\begin{figure*}[ht]
    \centering
    \includegraphics[width=1.0\linewidth]{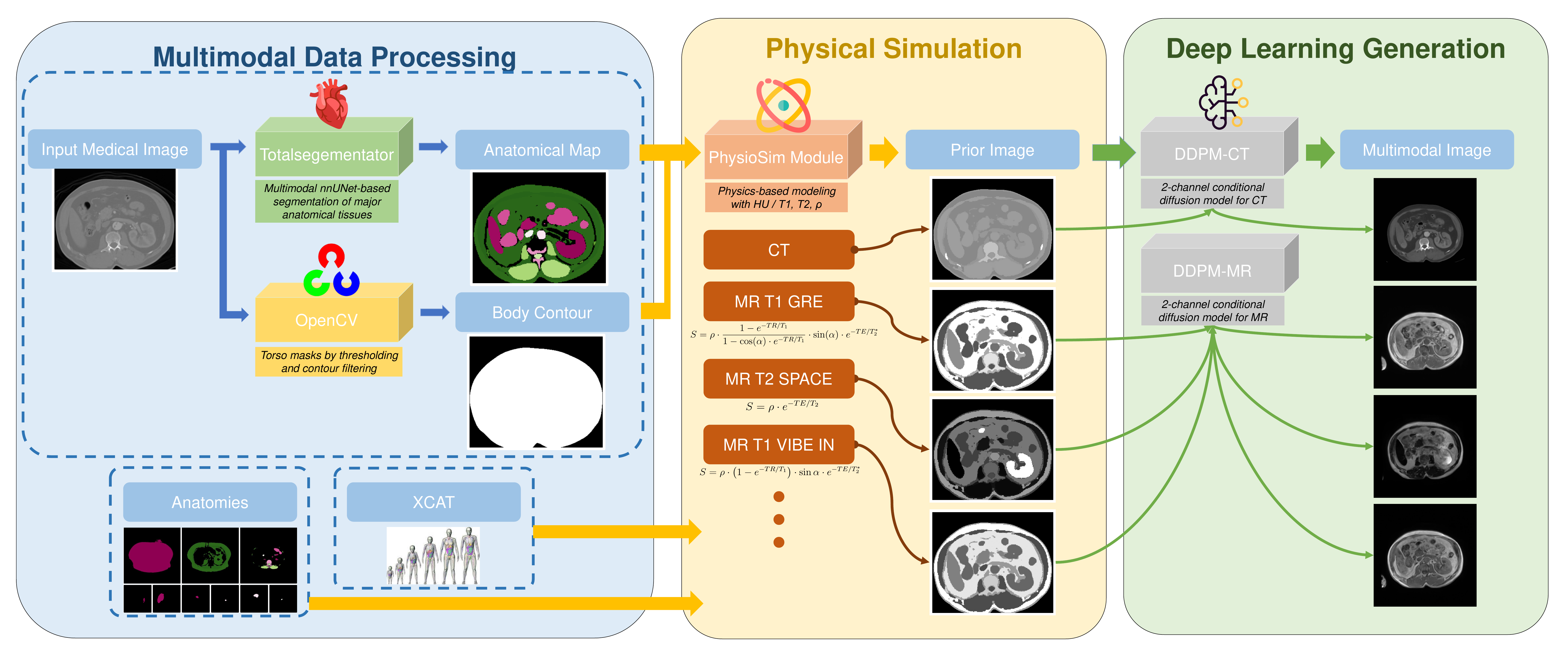}
    \caption{Illustration for the seg2med framework. The pipeline consists of three modular components. (1) Multimodal Data Processing: Anatomical maps are extracted from real CT/MR images, XCAT phantoms, or seg2med-assembled anatomies using TotalSegmentator and further refined with OpenCV-based body contour detection. (2) PhysioSynth Module: Sequence-specific physical priors are generated based on assigned HU or MR signal equations using tissue-dependent T1, T2, and $\rho$ values. (3) Deep Learning Generation: DDPM-CT and DDPM-MR take the anatomical prior as conditional input to generate structurally aligned, high-fidelity images across CT and multiple MR sequences.}
    \label{Figure_1}
\end{figure*}

The framework is depicted in Figure \ref{Figure_1}.

To construct anatomical priors for multimodal image generation, we developed a standardized preprocessing pipeline comprising two steps: (1) full-body segmentation via TotalSegmentator and (2) body contour extraction via OpenCV. These steps were applied separately to CT and MR volumes.

\subsubsection{Full-Organ and Tissue Segmentation}

We employed the TotalSegmentator toolkit to obtain comprehensive anatomical segmentations from CT and MR images. Specifically:
\begin{itemize}
    \item For CT images, both the \texttt{total} (full organ) and \texttt{tissue\_types} (subcutaneous fat, torso fat, skeletal muscle) modes were used to generate masks covering up to 117 anatomical classes.
    \item For MR images, due to differences in contrast and signal characteristics, we used \texttt{total\_mr} and \texttt{tissue\_types\_mr} modes to obtain MR-specific anatomical and tissue segmentations.
\end{itemize}
These masks were resampled to a consistent resolution of $1.0 \times 1.0$ mm$^2$ in the axial plane and stacked to form 3D volumes. Each segmentation was stored as a separate binary map for subsequent simulation and visualization.

\subsubsection{Body Contour Extraction and Fusion}

To isolate the patient's body region and eliminate irrelevant structures (e.g., scanner bed, arms), we applied contour detection techniques using OpenCV. The process involved:
\begin{enumerate}
    \item Thresholding the image to isolate the soft-tissue range.
    \item Using \texttt{cv2.findContours()} to detect outer edges.
    \item Applying \texttt{cv2.drawContours()} and \texttt{cv2.fillPoly()} to generate a solid binary mask of the body contour.
\end{enumerate}

The resulting contour mask was then combined with the anatomical segmentation mask to restrict organ priors to within the physical body region. This fusion ensured consistent spatial alignment across modalities and helped eliminate artifacts from overlapping segmentations. The fused masks served as the input to the PhysioSynth module for further simulation.

\subsection{PhysioSynth: Physics-Based Prior Simulation}

The PhysioSynth module generates modality-specific image priors by converting anatomical segmentation masks into imaging-like volumes using tissue-dependent physical parameters and sequence-specific signal equations.

\subsubsection{CT Simulation}
For CT priors, each organ class in the segmentation mask is assigned a representative Hounsfield Unit (HU) value based on established anatomical attenuation tables. These HU values simulate realistic grayscale intensity distributions for different tissues (e.g., soft tissue, bone, fat, air).

\subsubsection{MR Simulation}
For MRI priors, signal intensities are computed using analytical equations corresponding to the physical behavior of different MR sequences. For each organ, literature-derived T1, T2, and proton density ($\rho$) values are used, and standard sequence parameters (TR, TE, flip angle $\alpha$) are applied. The following equations were used to simulate different T1- and T2-weighted MR sequences:

\begin{itemize}
    \item \textbf{T1-weighted GRE:}
    \begin{equation}
        S = \rho \cdot \sin(\alpha) \cdot \frac{1 - e^{-TR/T1}}{1 - \cos(\alpha) \cdot e^{-TR/T1}}
    \end{equation}

    \item \textbf{T2-weighted SPACE:}
    \begin{equation}
        S = \rho \cdot e^{-TE/T2}
    \end{equation}

    \item \textbf{T1-weighted VIBE (in-phase and opposed-phase):}
    \begin{equation}
        S = \rho \cdot \sin(\alpha) \cdot \left( \frac{1 - e^{-TR/T1}}{1 - \cos(\alpha) \cdot e^{-TR/T1}} \right)
    \end{equation}
    Although the same signal equation is used for both in-phase and opposed-phase VIBE images, signal assignments are adjusted for fat–water mixed regions (e.g., liver, subcutaneous fat) to reflect phase interference effects.

    \item \textbf{T1-weighted Dixon VIBE (TRA in-phase and opposed-phase):}\\
    The same GRE-based equation is applied. To simulate the fat–water cancellation in opposed-phase images and enhancement in in-phase images, signal intensity scaling is applied to fat-containing organs.
\end{itemize}

In all simulations, organ-specific $T1$, $T2$, and $\rho$ values are assigned using lookup tables derived from published studies. The resulting simulated MR priors reflect both anatomical structure and sequence-specific contrast properties, providing a physically meaningful, interpretable prior for training the generative model.

\subsection{Conditional Diffusion Model with Anatomical Prior}

We adopt a conditional denoising diffusion probabilistic model (DDPM) \cite{3476-ddpm} for modality-specific image synthesis. Separate models are trained for CT and MR, each conditioned on anatomical priors derived from segmentation masks.

\subsubsection{Condition Injection via Mask Concatenation}

Let $x_t$ denote the noisy target image at time step $t$, and $y$ the anatomical prior image generated from PhysioSynth. The conditional noise prediction function becomes:

\begin{equation}
    \epsilon_\theta(x_t, t \mid y)
\end{equation}

To incorporate the anatomical prior, the input to the Diffusion-UNet is constructed by concatenating $x_t$ and $y$ as a two-channel input:

\begin{equation}
    z_t = \text{concat}(x_t, y) \in \mathbb{R}^{2 \times H \times W}
\end{equation}

The time step $t$ is encoded via a sinusoidal positional embedding $\mathbf{e}_t$, which is injected into all residual blocks of the Diffusion-UNet, following standard DDPM designs.

\subsubsection{Training Objective}

The training objective is to predict the Gaussian noise $\epsilon$ added to the clean image $x_0$, conditioned on the anatomical prior $y$. The simplified DDPM loss function is given by:

\begin{equation}
    \mathcal{L}_{\text{simple}} = \mathbb{E}_{x_0, \epsilon, t, y} \left[ \left\| \epsilon - \epsilon_\theta(x_t, t \mid y) \right\|^2 \right]
\end{equation}

where the noisy input is constructed as:

\begin{equation}
    x_t = \sqrt{\bar{\alpha}_t} \, x_0 + \sqrt{1 - \bar{\alpha}_t} \, \epsilon, \quad \epsilon \sim \mathcal{N}(0, I)
\end{equation}

By conditioning on anatomical priors, the model learns to generate structurally consistent medical images aligned with organ-specific spatial context.

\subsubsection{Training Strategy}

To train the DDPM models, we used pairs of anatomical segmentation priors (generated via PhysioSynth) and real images as supervision. During training, a clean target image $x_0$ was corrupted using the standard forward diffusion process to obtain the noisy version $x_t$, which served as the model input. The anatomical prior was concatenated with $x_t$ along the channel dimension, resulting in a two-channel input. 

We trained the model for 300 epochs using the Adam optimizer with a learning rate of $2 \times 10^{-4}$, $\beta_1 = 0.9$, and $\beta_2 = 0.999$. The noise schedule followed a linear $\beta_t$ schedule with 500 timesteps. All models were trained using a batch size of 8 on NVIDIA A6000 GPUs.

For comparative analysis, we evaluated the DDPM against three baseline generative models:
\begin{itemize}
    \item \textbf{Attention-UNet} \cite{3476-oktay2018attention}: a supervised image-to-image generator enhanced with spatial attention blocks.
    \item \textbf{Pix2Pix} \cite{3476-pix2pix}: a conditional GAN trained with paired inputs.
    \item \textbf{CycleGAN} \cite{zhuUnpairedImageImageTranslation2020}: an unpaired image translation model adapted to our task using pseudo-label pairs.
\end{itemize}

All baselines were trained on the same data splits with matched optimizer settings and evaluated using the same set of quantitative metrics (e.g., SSIM, MAE, FID). 

\subsection{Experiments}
\label{3476-ssec:experiments}
Overall, we have conducted five experiments: the generation of CT images, the generation of MR images, the conversion from digital phantoms to CT images including XCAT-to-CT conversion and synthetic-to-CT conversion, bi-directional image conversion between CT and MR modalities, and segmentation evaluation of synthetic CT images. 

\subsubsection{Experiment I}

In this experiment, we evaluated the performance of the proposed DDPM model on CT image synthesis (DDPM-CT) and compared it with three baseline generative models: Attention-UNet \cite{3476-oktay2018attention}, Pix2Pix \cite{3476-pix2pix}, and CycleGAN \cite{zhuUnpairedImageImageTranslation2020}.

The models were trained using two CT datasets: 170 pelvis CT images from SynthRAD Pelvis \cite{3476-synthrad} and 153 abdomen CT images from an internal dataset \cite{Anish_dataset}. Anatomical priors were generated as described before.

All models were trained using the same optimizer settings and batch sizes, and evaluated on three testing sets:
\begin{itemize}
    \item 10 unseen patients from SynthRAD Pelvis CT
    \item 10 unseen patients from Internal Abdomen CT
    \item All 48 patients from M2OLIE Abdomen CT
\end{itemize}

Evaluation metrics included SSIM, MAE, FSIM, and FID scores, comparing synthesized images with real CT ground truth. 

\subsubsection{Experiment II}

The second experiment assessed DDPM’s ability to synthesize MR images of varying sequences (DDPM-MR). Training data included:
\begin{itemize}
    \item 170 pelvis MR images from SynthRAD Pelvis, including 115 T1-weighted GRE and 55 T2-weighted SPACE sequences
    \item 81 T1-weighted MR images from M2OLIE Abdomen, including 37 VIBE in-phase, 37 opposed-phase, 3 Dixon in-phase, and 4 Dixon opposed-phase sequences
\end{itemize}

The model was tested on:
\begin{itemize}
    \item 40 unseen MR images from SynthRAD Pelvis
    \item 10 unseen MR images from M2OLIE Abdomen
\end{itemize}

\subsubsection{Experiment III}
The trained DDPM-CT was employed to generate CT images using masks from digital phantoms that lacked corresponding ground truth images. CT image generation was performed for 56 Extended Cardiac-Torso (XCAT) digital phantoms \cite{segars4DXCATPhantom2010}. Additionally, a composite digital phantom, createdby merging segmentations and body contour of multiple patients, was used to generate a CT image within our framework, showcasing the model's robustness and its potential for data augmentation.

\subsubsection{Experiment IV}
The image conversion between CT and MR modalities was performed by inputting CT masks into the DDPM-MR model and MR masks into the DDPM-CT model. 10 paired pelvis CT and images from the SynthRAD dataset were utilized in this experiment.
Due to differences in organ class indices between CT and MR segmentations in the Totalsegmentator toolkit, segmentations from the source modality were first converted to match the segmentation scheme of the target modality. These converted segmentations were then regularized using the body contour derived from the source image and assigned simulated values corresponding to the target modality. The resulting images were subsequently evaluated against the paired images of the target modality.

\subsubsection{Experiment V}
The synthetic abdomen CT images generated from segmentations of 10 patients from the internal dataset (Internal Abdomen) and 48 patients from the M2OLIE Abdomen CT/MR dataset were further segmented using the TotalSegmentator toolkit and evaluated by comparing with the original input segmentations.

\subsection{Evaluation Metrics}
\label{3476-ssec:eval}
In Experiment I, II and IV, the synthetic CT and MR images (sCT and sMR) were evaluated against the original CT and MR images using several image quality assessment metrics, including SSIM, Mean Absolute Error (MAE), Peak Signal-to-Noise Ratio (PSNR).
SSIM evaluates the structural similarity between the generated image and the ground truth image, taking into account factors like luminance, contrast, and structure. SSIM values range from -1 to 1, with 1 indicating a perfect similarity.
MAE measures the average absolute difference between the generated image and the ground truth image. A lower MAE indicates that the generated image is closer to the ground truth image in terms of pixel-level differences. 
PSNR (Peak Signal-to-Noise Ratio) measures the ratio between the maximum possible power of a signal (the pixel values of the original image) and the power of the noise (the difference between the original and generated image). A higher PSNR value is generally associated with better image quality.
Additionally, the consistency between the histograms of synthetic images and ground truth images was evaluated using Histogram Correlation Coefficient (HistCC). 

To determine the statistical significance of the improvements in the SSIM, MAE, and PSNR, we performed statistical hypothesis tests. We compared the DDPM-CT with each other model and applied a Mann-Whitney U test as the metrics are non-normally distributed. Our null hypothesis states that the DDPM did not lead to improvements in either metric. The null hypothesis is rejected if p \textless 0.05.

The Frèchet Inception Distance (FID) \cite{heuselGANsTrainedTwo2018} is a widely used metric for evaluating image synthesis, quantifying the similarity between two distributions—typically real and generated images. Lower FID scores indicate greater similarity, with values approaching zero reflecting better performance. In this study, we employed the standard implementation using the Inception-V3 model \cite{szegedyRethinkingInceptionArchitecture2015} to compute the FID between the synthesized images and the ground-truth images. Since the synthesized medical images are single-channel and have a broader pixel value range compared to RGB images, we adapted them by duplicating to three channels and normalizing the pixel values to the range of 0–1. A P-test is not required for FID, as it inherently quantifies the differences between the distributions of real and generated images.

Feature Similarity Index (FSIM) is suitable for evaluating synthetic CT images generated from XCAT-derived masks in experiment III, as it does not rely on ground truth for comparison. It is a feature-based metric, evaluates the perceptual similarity of images by focusing on critical low-level features such as edges and textures, making it especially effective for assessing structural detail \cite{linzhangFSIMFeatureSimilarity2011}. 

The Dice coefficient was used to quantify the overlap between segmented regions in synthetic and target images in experiment V, offering insight into organ-level accuracy.

\section{Results}
\label{3476-sec:results}

\subsection{Experiment I: Generation of CT Images}
\label{3476-ssec:ExperimentI}

Table \ref{table_ct_four_models} provides the quantitative results for CT image generation achieved by each implemented model. The models were trained using the training datasets from the SynthRAD Pelvis and the Internal Abdomen, while the M2OLIE Abdomen consisted of "unseen" patients for evaluation. The results indicate that, although DDPM-CT shows no significant advantage over other models in validation on the SynthRAD Pelvis and the Internal Abdomen, it demonstrates a clear improvement for the M2OLIE Abdomen. Specifically, DDPM-CT achieves the highest SSIM of \(0.93 \pm 0.02\), the lowest MAE of \(40.05 \pm 13.15\), a relatively high PSNR of \(26.16 \pm 1.62\), and the lowest standard deviation across all metrics, indicating its superior and more stable performance compared to other models. 
Overall, DDPM-CT achieves the highest SSIM of \(0.94 \pm 0.02\), the lowest MAE of \(36.79 \pm 24.34\), the lowest FID of 3.62, although the highest PSNR is observed with UNet (\(28.05 \pm 4.12\)).

Table \ref{table_p_value} presents the results of the significance tests, demonstrating that DDPM-CT significantly outperformed other models in SSIM and MAE metrics (p \textless 0.05) but performed significantly worse than other models in terms of PSNR (p \textless 0.05).

\begin{figure*}[htb]
    \centering
    \includegraphics[width=0.8\linewidth]{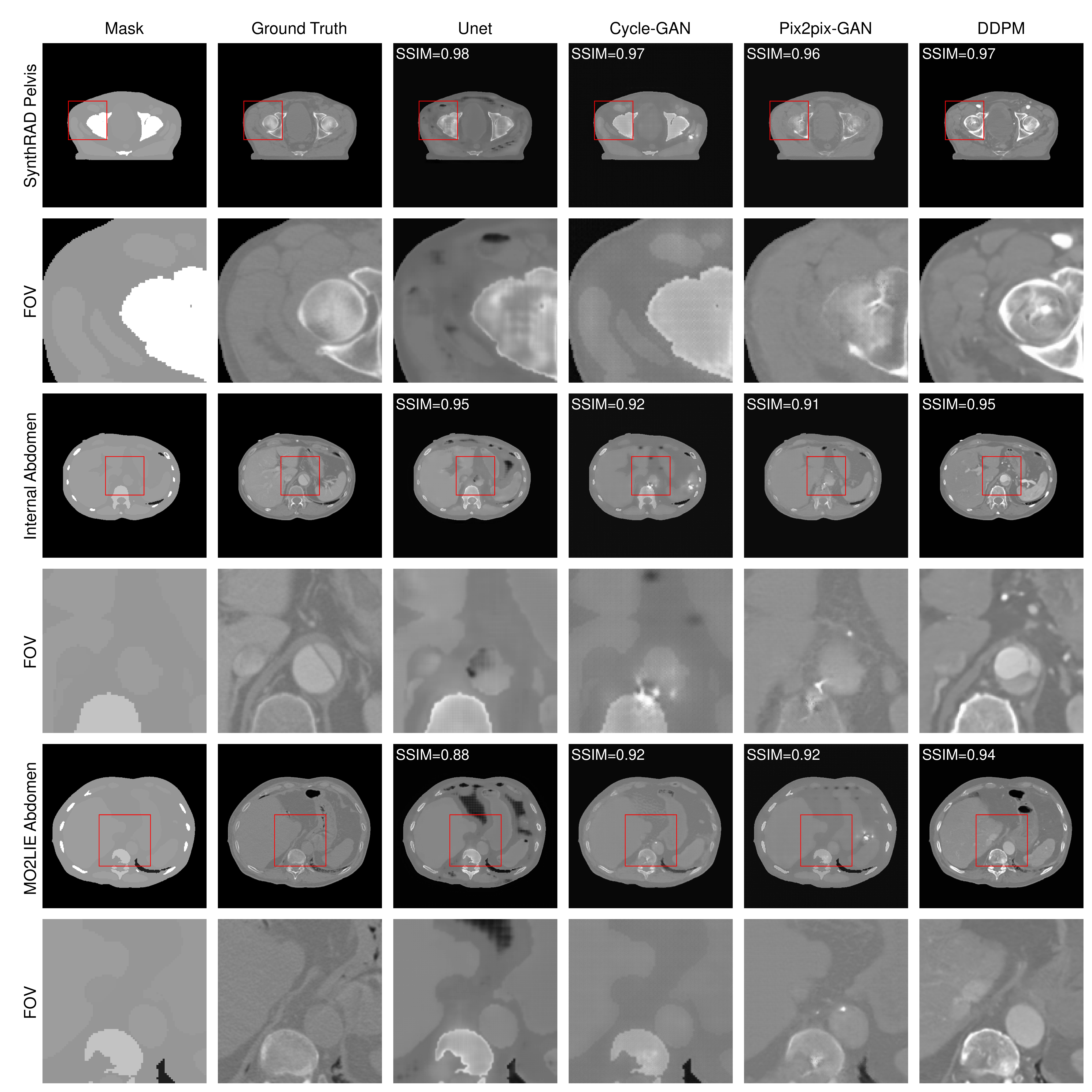}
    \caption{Qualitative comparison of synthetic CT images generated by four models across the three datasets. This figure presents one representative image slice per dataset, illustrating the outputs of the four models: UNet, CycleGAN, pix2pix, and DDPM. The first column displays the input mask, followed by the ground truth image. Subsequent columns show synthetic images generated by each model. Key metrics (SSIM, PSNR, and MAE) are provided for quantitative comparison. Focused fields of view (FOVs) within the red boxes highlight regions of interest to enhance visual comparison of structural details and model performance.}
    \label{Figure_2}
\end{figure*}

Representative examples of three CT image slices from a patient in each dataset are illustrated in Figure \ref{Figure_2}, with SSIM noted on each synthetic CT image. These examples highlight DDPM's ability to achieve strong quantitative metrics while preserving anatomical details such as the aorta, kidney, and pancreas, as evident in the field of view (FOV) boxes.

\begin{figure*}[htb]
    \centering
    \includegraphics[width=0.9\linewidth]{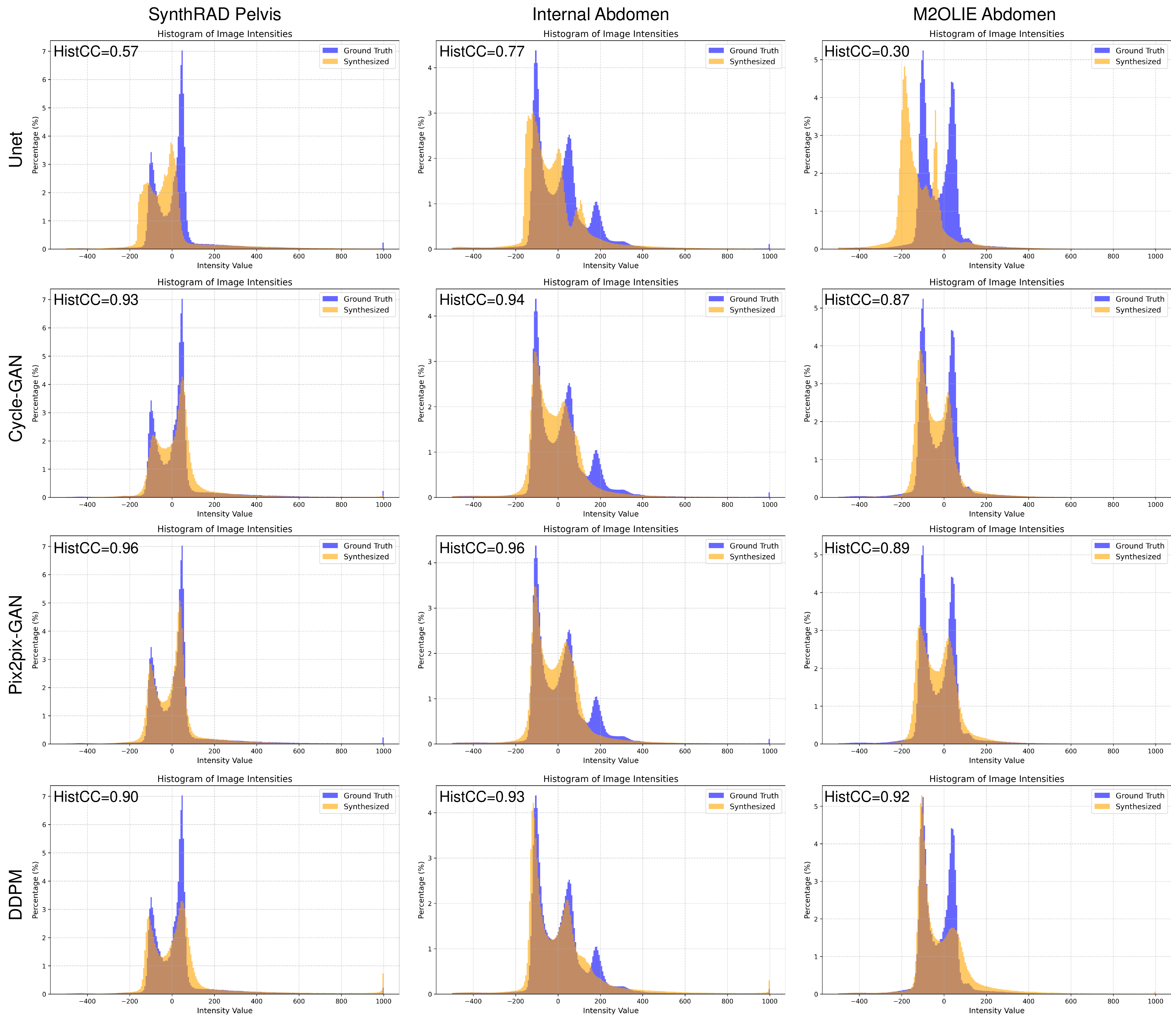}
    \caption{Comparison of image intensities distribution across models and datasets. Histograms of image intensities for ground truth (blue) and synthesized (orange) images corresponding to the image slices are shown. Histograms are provided for three datasets and four deep learning models: UNet, CycleGAN, pix2pix, and DDPM. The Histogram Correlation Coefficient (HistCC) values, representing the similarity between the intensity distributions of synthesized and ground truth images, are calculated and displayed for each case. These visualizations highlight the alignment of intensity distributions achieved by each model within the respective datasets.}
    \label{Figure_3}
\end{figure*}

An analysis of the histograms comparing synthetic CT images from each model with the ground-truth CT images of the same patients shows that the DDPM-generated images closely match the real CT images in terms of intensity distribution, as illustrated in Figure \ref{Figure_3}. For the example patient, the histogram correlation coefficients (HistCCs) of the DDPM-generated images from all three datasets exceed 0.90.

\begin{figure*}[htb]
    \centering
    \includegraphics[width=1.0\linewidth]{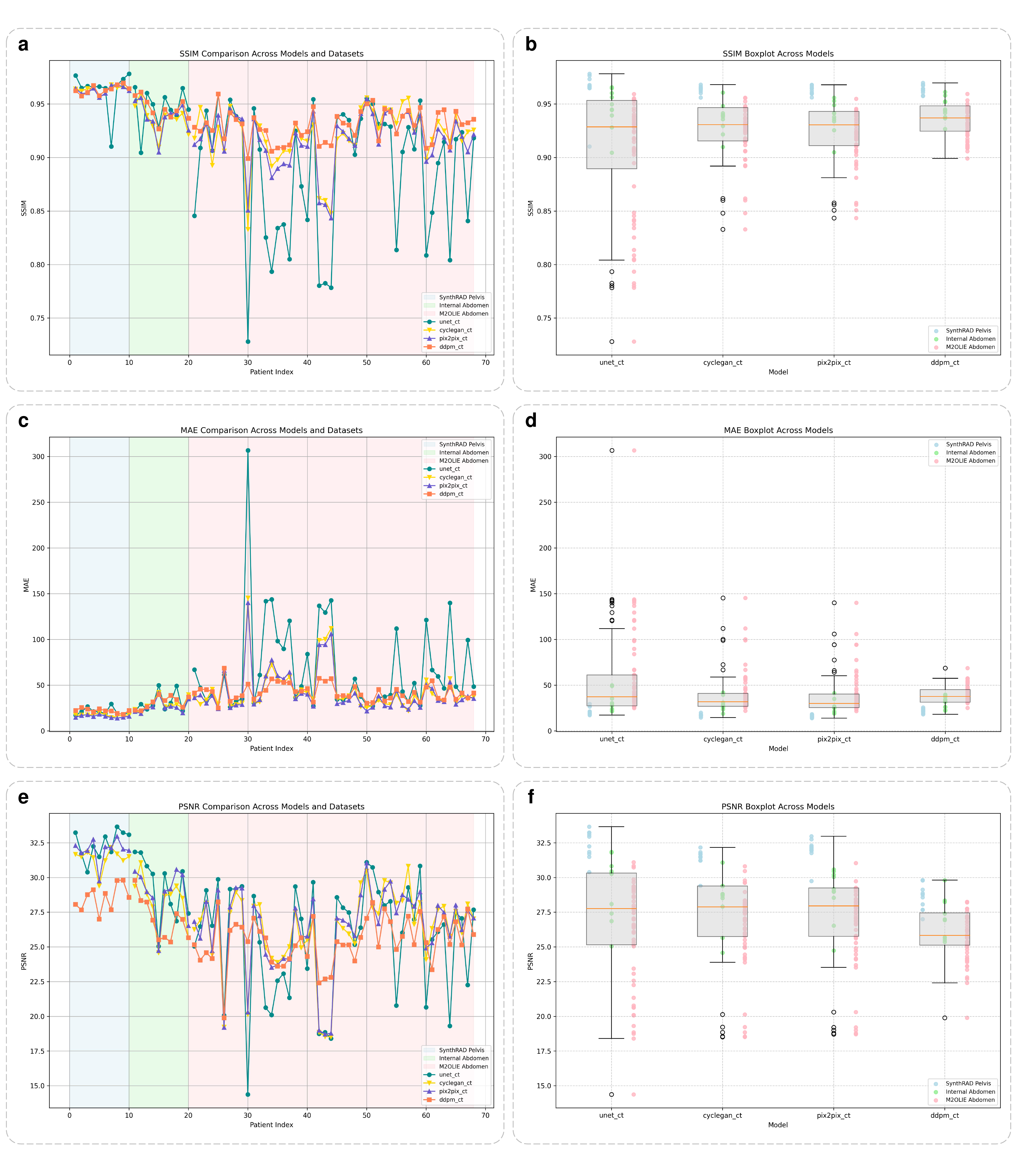}
    \caption{Comparison of metric across models and datasets by line charts and box plots. This figure illustrates the performance of the four deep learning models—UNet, CycleGAN, pix2pix, and DDPM—on 68 test patients across three distinct datasets. The datasets are visually separated by the different background colors in line charts and the different point colors in box plots to facilitate comparative analysis, with blue standing for the SynthRAD Pelvis dataset, green standing for the Internal Abdomen dataset, and pink standing for the M2OLIE Abdomen-dataset. Figures (a), (c), and (d) present line charts for SSIM, MAE, and PSNR, with the patient index on the x-axis. Figures (b), (d), and (f) display box plots for SSIM, MAE, and PSNR, with the x-axis representing the different models}.
    \label{Figure_4}
\end{figure*}

To provide a comprehensive comparison of the four models, Figure \ref{Figure_4} presents line charts and box plots of the metrics for synthetic CT images of test patients across all three datasets. In line charts, the x-axis represents the patient index, and the y-axis indicates the mean metric value for the generated CT images of each patient. The three datasets are distinguished by background colors. 
In the box plots, the x-axis represents the four models, and the y-axis shows the metric values. Additionally, the exact values for all evaluated patients are displayed as points, with each dataset distinguished by a unique point color. 
The line charts and box plots show that DDPM-CT consistently delivers stable and excellent performance across all datasets, while the other models exhibit poorer performance for some patients, particularly in the new dataset (e.g., patients 41, 42, and 43). This observation is further supported by DDPM-CT's lowest standard deviation across all metrics in Table \ref{table_ct_four_models}.

\begin{table}[t!]
\centering
\caption{Quantitative evaluation of the four models on the three test CT datasets. The SynthRAD Pelvis includes ten unseen patients from the synthrad dataset \cite{3476-synthrad}. The Internal Abdomen includes ten unseen patients from an internal abdomen-CT dataset \cite{Anish_dataset}. The M2OLIE Abdomen is an abdomen CT dataset from the M2OLIE project, which is not included in training and is an independent dataset \cite{Anika_dataset}. Values are presented as mean$\pm$standard deviation. Better results are indicated by higher SSIM and PSNR values, along with lower MAE and FID values.}
\vspace*{-1em}

\begin{subtable} {.5\textwidth}
\scriptsize
\caption{SSIM}
\begin{tabular*}{\textwidth}{c@{\extracolsep\fill}cccc}
\toprule
mean$\pm$std &UNet&CycleGAN&pix2pix&DDPM-CT \\ 
\midrule
SynthRAD Pelvis&0.96$\pm$0.02&0.96$\pm$0.01&0.96$\pm$0.01&0.96$\pm$0.01\\ 
Internal Abdomen&0.95$\pm$0.03&0.94$\pm$0.02&0.94$\pm$0.02&0.95$\pm$0.02\\ 
M2OLIE Abdomen&0.89$\pm$0.06&0.92$\pm$0.03&0.92$\pm$0.03&0.93$\pm$0.02\\ 
overall &0.92$\pm$0.05&0.93$\pm$0.04&0.93$\pm$0.04&0.94$\pm$0.02\\ 
\bottomrule
\end{tabular*}
\label{subtable1}
\end{subtable}

\vspace{1em}

\begin{subtable}{.5\textwidth}
\scriptsize
\caption{PSNR}
\begin{tabular*}{\textwidth}{c@{\extracolsep\fill}cccc}
\toprule
mean$\pm$std &UNet&CycleGAN&pix2pix&DDPM-CT \\ 
\midrule
SynthRAD Pelvis&32.36$\pm$1.79&31.39$\pm$1.45&32.00$\pm$2.28&30.00$\pm$2.60 \\ 
Internal Abdomen&29.29$\pm$3.13&28.28$\pm$2.66&28.82$\pm$2.72&28.00$\pm$2.51 \\ 
M2OLIE Abdomen&25.98$\pm$3.88&26.37$\pm$3.10&26.49$\pm$3.03&26.16$\pm$1.62 \\ 
overall &28.05$\pm$4.12&27.90$\pm$3.71&27.97$\pm$3.62&27.58$\pm$3.48\\ 
\bottomrule
\end{tabular*}
\label{subtable2}
\end{subtable}

\vspace{1em}

\begin{subtable}{.5\textwidth}
\scriptsize
\caption{MAE}
\begin{tabular*}{\textwidth}{c@{\extracolsep\fill}cccc}
\toprule
mean$\pm$std &UNet&CycleGAN&pix2pix&DDPM \\ 
\midrule
SynthRAD Pelvis&21.35$\pm$5.08&17.36$\pm$2.25&15.98$\pm$3.11&19.60$\pm$4.94 \\ 
Internal Abdomen&31.33$\pm$12.82&29.10$\pm$9.92&26.61$\pm$9.32&28.95$\pm$10.16 \\ 
M2OLIE Abdomen&61.56$\pm$45.49&42.26$\pm$21.53&40.69$\pm$20.89&40.05$\pm$13.15 \\ 
overall &45.30$\pm$37.44&39.70$\pm$30.16&37.22$\pm$27.05&36.79$\pm$24.34\\
\bottomrule
\end{tabular*}
\label{subtable3}
\end{subtable}

\vspace{1em}

\begin{subtable}{.5\textwidth}
\scriptsize
\caption{FID}
\begin{tabular*}{\textwidth}{c@{\extracolsep\fill}cccc}
\toprule
 &UNet&CycleGAN&pix2pix&DDPM \\ 
\midrule
SynthRAD Pelvis & 8.56 & 7.62 & 3.48 & 4.46 \\
Internal Abdomen & 12.78 & 10.60 & 6.74 & 5.30 \\
M2OLIE Abdomen & 15.35 & 14.34 & 10.14 & 5.01 \\
overall & 11.26 & 9.98 & 6.25 & 3.62 \\
\bottomrule
\end{tabular*}
\label{subtable4}
\end{subtable}

\label{table_ct_four_models}
\end{table}

\begin{table}[t!]
    \centering
    \caption{P-values from the Mann-Whitney U test (p\textless0.05) assessing the significance of differences in SSIM, PSNR, and MAE metrics between DDPM and other models}
    \begin{tabular}{lccc}
    \toprule
    Model & \textbf{SSIM} & \textbf{PSNR} & \textbf{MAE} \\
    \midrule
    DDPM vs. UNet & \textless0.00001 & \textless0.00001 & \textless0.00001 \\
    DDPM vs. CycleGAN & \textless0.00001 & \textless0.00001 & 0.00710 \\
    DDPM vs. pix2pix & \textless0.00001 & \textless0.00001 & \textless0.00001 \\
    \bottomrule
    \end{tabular}
\label{table_p_value}
\end{table}

\subsection{Experiment II: Generation of MR Images}
\label{3476-ssec:ExperimentII}
\begin{figure*}[ht]
    \centering
    \includegraphics[width=0.8\linewidth]{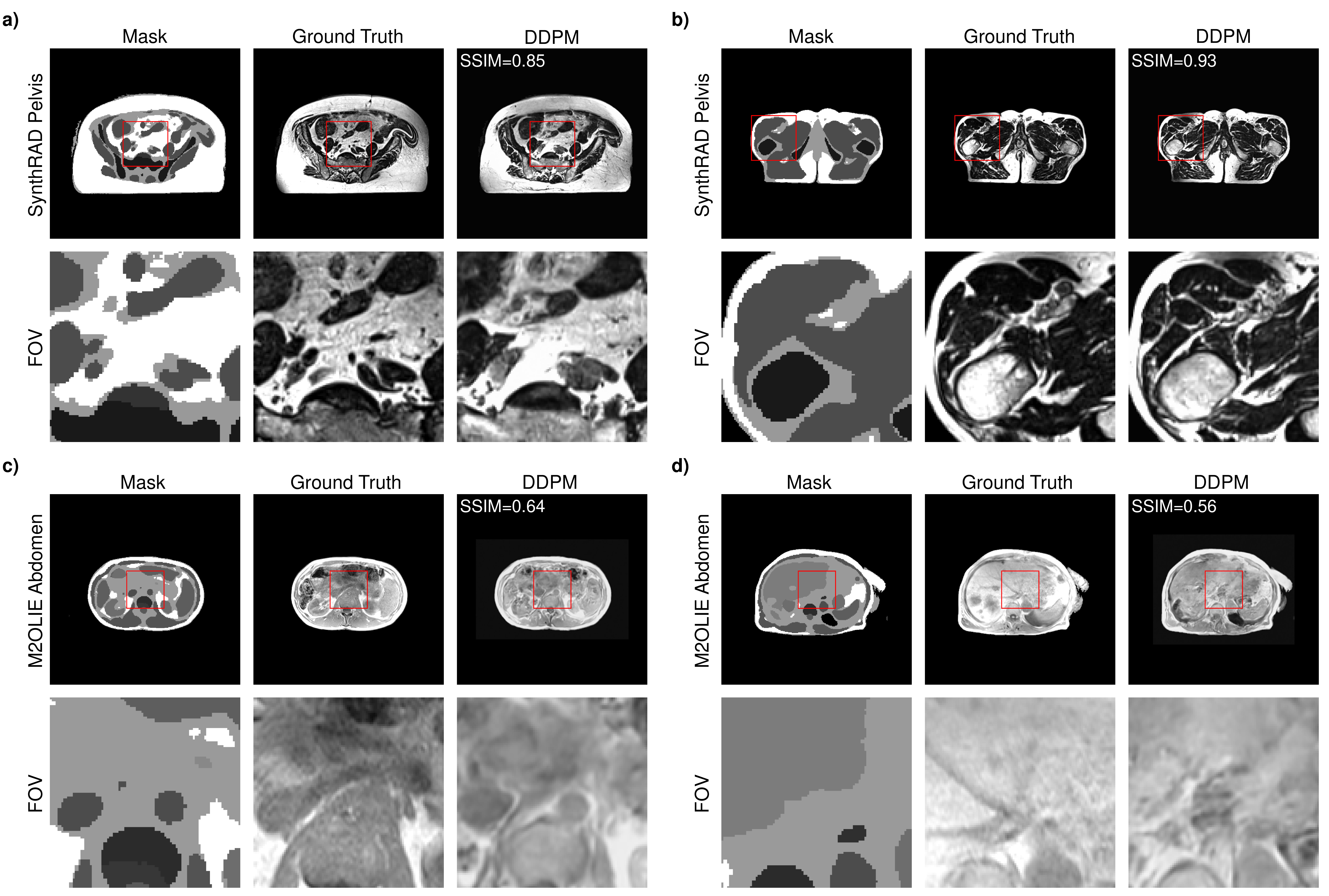}
    \caption{Qualitative comparison of synthetic MR images generated by DDPM across two MR datasets. This figure presents representative slices of synthetic MR images for two datasets, showcasing the input masks (left column) and the corresponding outputs from the DDPM model (right column). Key quantitative metrics, including SSIM, PSNR, and MAE, are displayed to evaluate the model's performance. The rows highlight different anatomical structures, with red boxes marking fields of view (FOVs) for detailed comparison. Enlarged views of the FOVs are provided to illustrate structural and textural fidelity between the input masks and the generated MR images.}
    \label{Figure_5}
\end{figure*}

Two MR datasets were utilized for the second experiment: the pelvis MR image dataset from Task 1 of the SynthRad2023 dataset (SynthRAD Pelvis) and the M2OLIE abdomen MR dataset (M2OLIE Abdomen). The test results were summarized in Table \ref{table_mr}, with the value range of the generated MR images normalized to [0, 255]. The pelvis MR images generated by our framework achieved a mean SSIM close to 0.90, demonstrating high image quality. In contrast, the generated abdomen MR images for M2OLIE Abdomen exhibited lower quality, with a mean SSIM of $0.56 \pm 0.07$. This disparity is attributed to the larger dataset size and superior image quality of the SynthRAD Pelvis compared to the M2OLIE Abdomen. As shown in Figure \ref{Figure_5}, the synthesized pelvis MR images generated by DDPM display well-preserved image details, particularly in skeletal and muscle tissue.

\begin{table}[t!]
    \scriptsize
    \centering
    \caption{Metrics for assessing quality of synthetic MR images}
    \begin{tabular}{lcccc}
    \toprule
     & \textbf{SSIM} & \textbf{PSNR} & \textbf{MAE} & \textbf{FID} \\
    \midrule
    SynthRAD Pelvis & $0.89 \pm 0.04$ & $20.56 \pm 3.52$ & $9.70 \pm 4.68$ & 4.64\\
    M2OLIE Abdomen & $0.56 \pm 0.07$ & $22.67 \pm 6.05$ & $ 17.75 \pm 1.68$ & 36.66\\
    \bottomrule
    \end{tabular}
\label{table_mr}
\end{table}

\subsection{Experiment III: Conversion from Digital Phantoms to CT Images}
\label{3476-ssec:ExperimentIII}

\begin{figure*}[htb]
    \centering
    \includegraphics[width=0.8\linewidth]{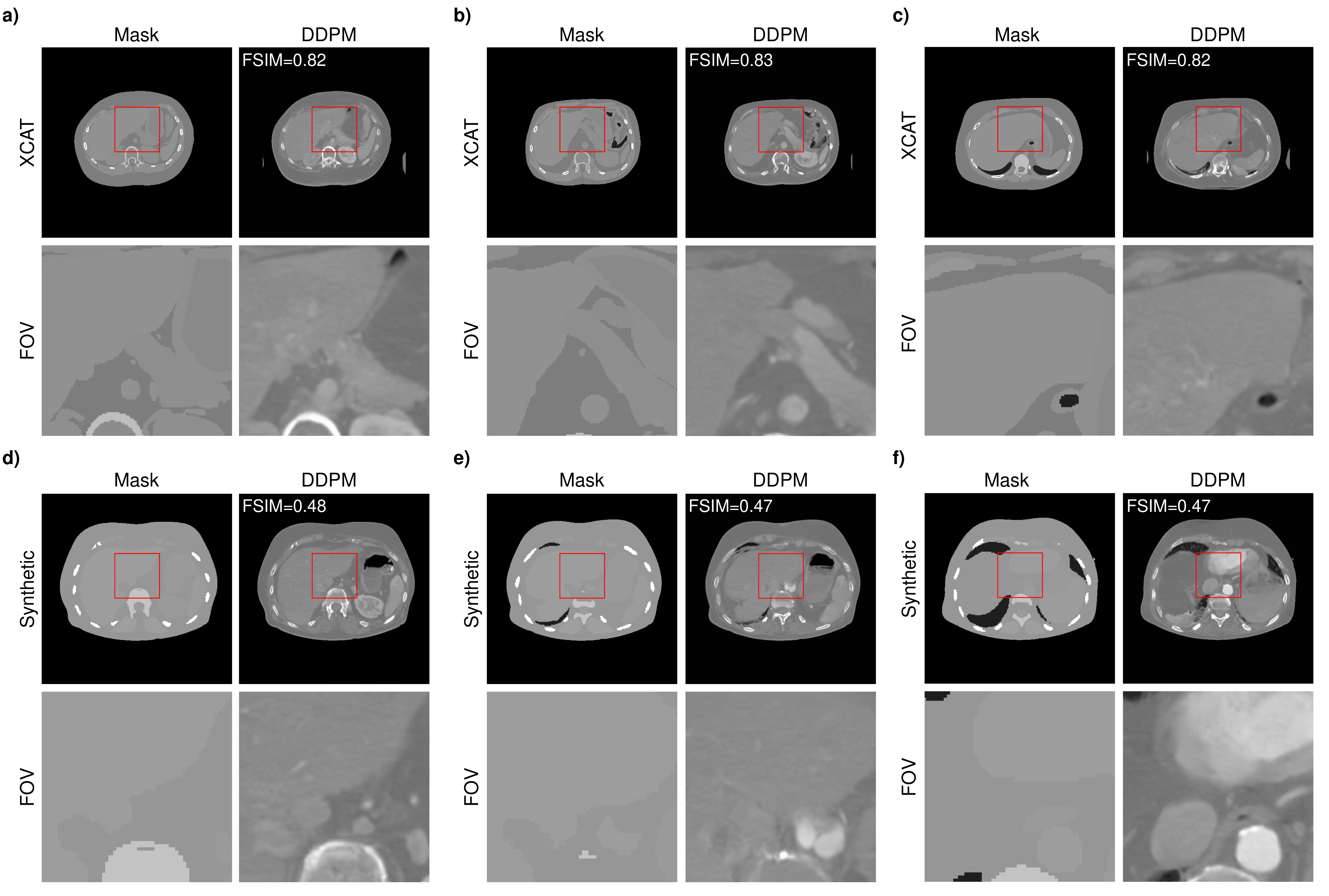}
    \caption{Qualitative Assessment of Synthetic CT Images Generated by DDPM for XCAT and Synthetic Data. This figure presents representative slices of XCAT-masks and synthetic CT images, showcasing the input masks (left column) and the corresponding outputs generated by the DDPM model (right column). Feature Similarity Index (FSIM) is displayed for each example to evaluate the structural and intensity similarities between the generated images and the real CT images. Red boxes indicate fields of view (FOV), with magnified views provided below each slice to highlight fine details and differences in texture and intensity. This assessment emphasizes the model's performance on digital phantoms without ground truth.}
    \label{Figure_6}
\end{figure*}

\begin{figure*}[htb]
    \centering
    \includegraphics[width=0.8\linewidth]{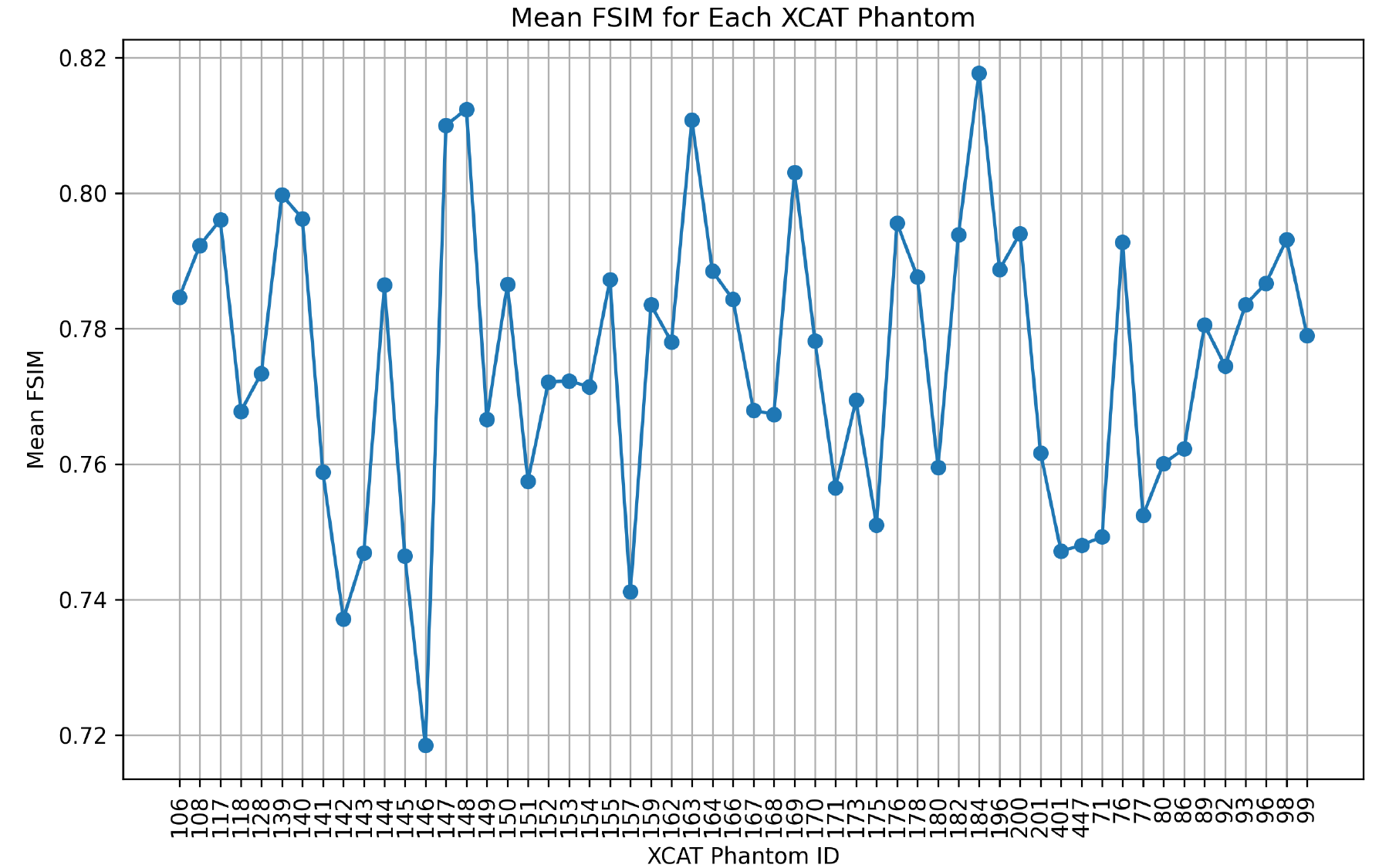}
    \caption{Mean FSIM for Each XCAT Phantom. The figure presents the mean Feature Similarity Index Measure (FSIM) for synthetic images generated for each XCAT phantom. FSIM quantifies the perceptual similarity of the images by evaluating feature consistency, such as edges and textures. Most phantoms exhibit FSIM values exceeding 0.75, indicating high-quality synthetic image generation with well-preserved structural details. Variations in FSIM across phantoms may arise from differences in anatomical features and image complexity inherent to each phantom.}
    \label{Figure_7}
\end{figure*}

Table \ref{tab:fsim} presents the mean FSIM values for synthesized CT images generated from 56 XCAT digital phantoms and a digital phantom created by merging segmentations and body contours from multiple patients. The CT images generated from segmentations of real patient images in Experiment I achieved a relatively high mean FSIM of \(0.82 \pm 0.08\), while the CT images synthesized from XCAT digital phantoms exhibited a comparable FSIM of \(0.78 \pm 0.04\), indicating a high degree of structural consistency with real images. In contrast, the CT image generated from the merged-segmentation digital phantom showed a lower FSIM of \(0.45 \pm 0.02\). 
Figure \ref{Figure_6} illustrates a qualitative assessment of synthetic CT images generated by DDPM for XCAT and synthetic digital phantoms. Three image slices are selected from different XCAT models (Model89, Model106, Model140), while the bottom three slices are derived from a single synthetic digital phantom created by merging segmentations from three different patients. As shown in the figure, the synthetic CT images generated from XCAT phantoms preserve realistic anatomical structures, whereas the image synthesized from the synthetic digital phantom does not. This qualitative observation aligns with the FSIM values displayed on the upleft side of the figures.
Figure \ref{Figure_7} provides an overview of the FSIM values for the CT images synthesized from the 56 XCAT digital phantoms. Most phantoms achieved FSIM values above 0.75, reflecting high-quality image synthesis with well-preserved structural details. The observed variation in FSIM values may be attributed to differences in anatomical complexity and feature representation among the phantoms.

\begin{table}[t!]
    \centering
    \caption{FSIM for assessing quality of synthetic images from XCAT digital phantoms and synthetic digital phantoms created by segmentations from multiple patients}
    \begin{tabular}{lccc}
    \toprule
     & Synthetic CT & XCAT CT & Synthetic Patient CT \\
      & (Experiment I) & & \\
    \midrule
    FSIM & $0.82 \pm 0.08$ & $0.78 \pm 0.04$ & $0.45 \pm 0.02$ \\
    \bottomrule
    \end{tabular}
\label{tab:fsim}
\end{table}

\subsection{Experiment IV: CT-MR Image Conversion}
\label{3476-ssec:ExperimentIV}
\begin{figure*}[htb]
    \centering
    \includegraphics[width=0.8\linewidth]{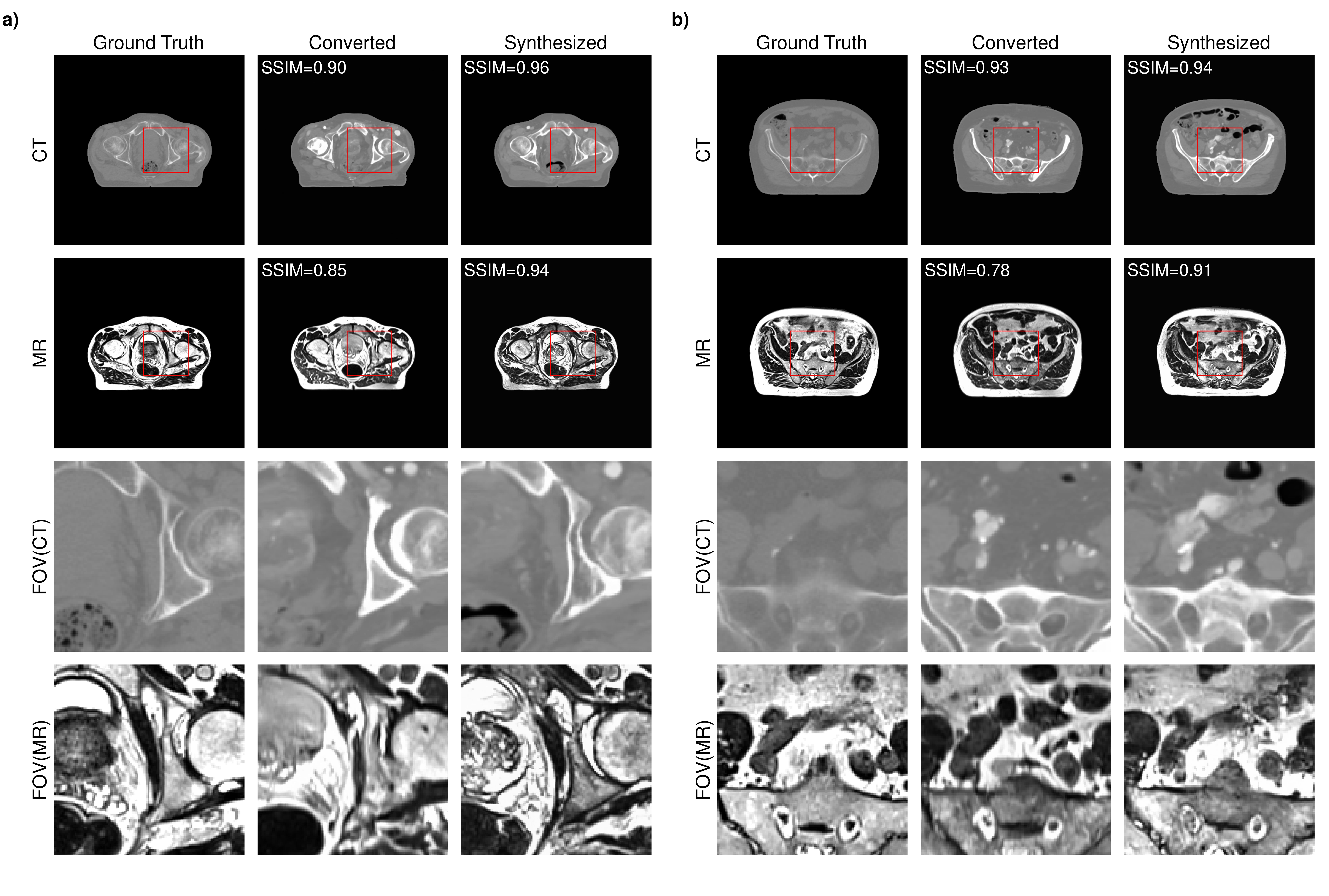}
    \caption{Qualitative and Quantitative Comparison of Ground Truth, Converted, and Synthesized Images for CT and MR Modalities. This figure presents representative image slices from CT (top row) and MR (bottom row) modalities, comparing the ground truth images, converted images, and synthesized images from Experiment I and II. The converted images are generated using the masks from the opposite modality. Specifically, converted CT images are produced using masks from the MR modality, and vice versa. Each modality includes two sets of comparisons, highlighting Fields of Views (FOVs) marked by red boxes. Quantitative metrics, including SSIM, PSNR, and MAE, are provided for each example to evaluate structural fidelity, intensity consistency, and overall error. Enlarged views of the FOVs are shown below each row to emphasize fine structural and textural details, facilitating a detailed comparison between ground truth, converted, and synthesized outputs.}
    \label{Figure_8}
\end{figure*}

The dataset comprising 10 paired pelvis CT and MR images was utilized in this experiment. Table \ref{table_ct2mr} summarizes the mean SSIM, PSNR, and MAE values for the 10 converted CT and MR images. The results indicate that the converted CT images achieve a high SSIM of $0.91\pm0.03$, demonstrating their high structural fidelity. In contrast, the converted MR images exhibit lower quality, with a relatively low SSIM of $0.77\pm0.04$ and a PSNR of $15.53\pm2.02$. Figure \ref{Figure_8} displays the converted images alongside the ground truth images of the target modality and the synthesized images from Experiment I and Experiment II for the same patient. Examples from two patients are included in the figure. The comparison shows that the converted CT images effectively preserve critical anatomical structures, while the converted MR images maintain good quality and similarity to ground truth but lack detail in certain tissues.

\begin{table}[t!]
    \centering
    \caption{Metrics for assessing quality of image conversion between CT and MR images.}
    \begin{tabular}{lccc}
    \toprule
     & \textbf{SSIM} & \textbf{PSNR} & \textbf{MAE} \\
    \midrule
    MR-to-CT & $0.91 \pm 0.03$ & $22.76 \pm 2.35$ & $47.14 \pm 30.38$ \\
    CT-to-MR & $0.77 \pm 0.04$ & $15.53 \pm 1.47$ & $ 16.32 \pm 4.66$ \\
    \bottomrule
    \end{tabular}
\label{table_ct2mr}
\end{table}

\subsection{Experiment V: Segmentation Evaluation of Synthetic CT Images}
\label{3476-ssec:ExperimentV}
The synthetic abdomen CT images generated from the segmentations of 10 patients from the internal dataset (Internal Abdomen) and 48 patients from the M2OLIE Abdomen dataset (M2OLIE Abdomen) were further segmented using the Totalsegmentator toolkit's \textit{total} task. The mean Dice coefficients between the segmentations of each abdominal organ from the synthetic and ground truth images of these 58 patients are presented in Table \ref{tab:mean_dice_coefficients}. The results indicate that 11 of the 59 organs achieved a mean Dice coefficient greater than 0.90, demonstrating high consistency with the original CT segmentations. These organs include Autochthon Left, Autochthon Right, Kidney Left, Kidney Right, Liver, Rib Left 11, Scapula Right, Spinal Cord, Vertebrae L3, Vertebrae T10, and Vertebrae T11. Additionally, 34 of the 59 organs achieved a mean Dice coefficient exceeding 0.80.

\begin{figure*}[htb]
    \centering
    \includegraphics[width=0.8\linewidth]{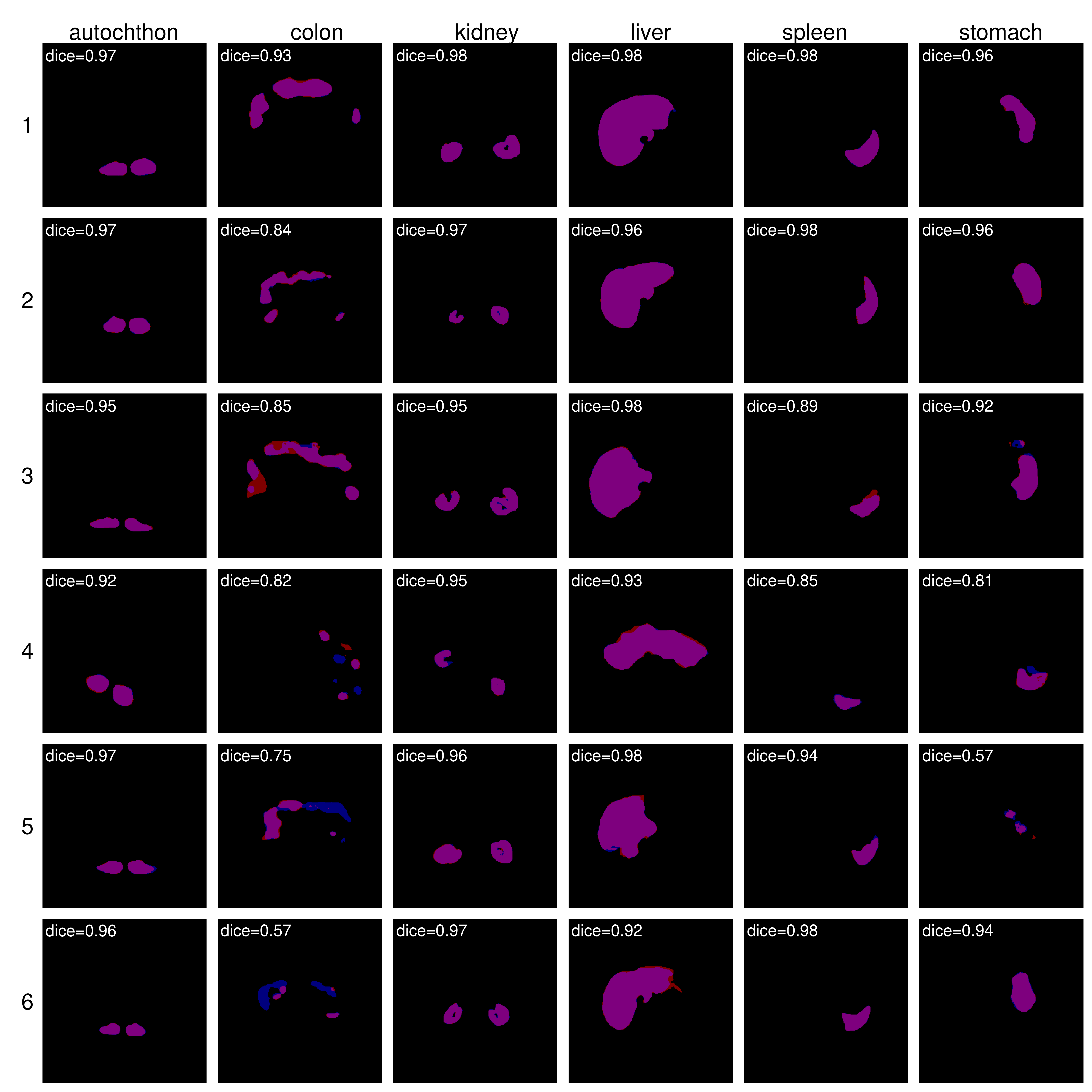}
    \caption{Organ segmentation results with dice scores across different organs and cases. This figure displays the segmentation overlaps for six organs (autochthon, colon, kidney, liver, spleen, and stomach) across six cases. The red regions represent segmentations from the ground truth images. The blue region represent the segmentations from the synthetic images. The purple regions represent the overlap, while non-overlapping areas indicate segmentation discrepancies. The Dice similarity coefficient (Dice) is calculated for each organ and displayed above each panel. The results demonstrate the consistency between segmentations derived from synthetic CT images and those from ground truth images, with higher Dice scores indicating better agreement.}
    \label{Figure_9}
\end{figure*}

Figure \ref{Figure_9} provides the organ segmentation results for six different organs across six patient cases, with the corresponding Dice scores annotated on each image. The figure illustrates that the segmentations of autochthon (left and right), kidney, and liver consistently achieve high Dice coefficients above 0.90 for all six patients. The segmentation of the colon in one patient also exceeds 0.90, while the segmentations of the spleen and stomach exceed 0.90 in four of the six cases.

\begin{figure*}[htb]
    \centering
    \includegraphics[width=1.0\linewidth]{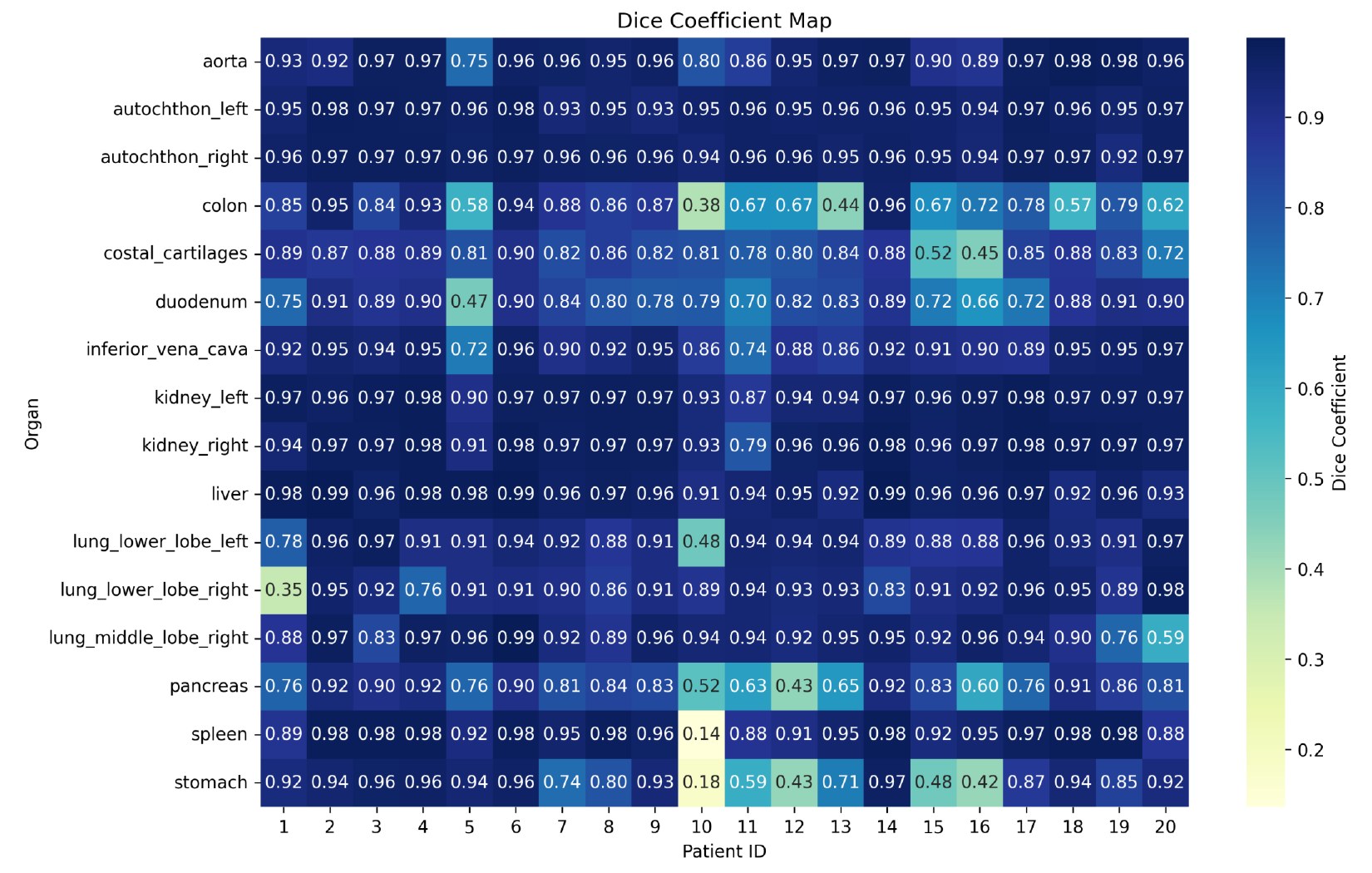}
    \caption{Dice coefficient keatmap for organ segmentation across patients. This heatmap illustrates the Dice similarity coefficients for organ segmentation across 20 patients and 16 abdominal organs, from aorta to stomach. Each cell represents the Dice coefficient for a specific organ and patient, with higher values (darker blue) indicating better segmentation overlap between segmentations derived from synthetic CT images and those from ground truth images. The color gradient highlights performance variability, with lighter shades revealing lower segmentation accuracy. This visualization provides a comprehensive overview of the consistency between segmentations derived from synthetic CT images and those from ground truth images, identifying organs or cases with higher or lower agreement.}
    \label{Figure_10}
\end{figure*}

Figure \ref{Figure_10} depicts a heatmap of Dice coefficients for 16 organ segmentations across 20 patients, covering a total of 320 cases. Higher values (darker blue) represent better segmentation overlap between synthetic image segmentations and ground truth segmentations. The heatmap reveals that 223 out of the 320 cases (69.69\%) achieved a Dice coefficient greater than 0.90, highlighting the strong agreement between synthetic and ground truth segmentations for most cases.

\begin{table*}[ht]
\centering
\caption{Mean Dice coefficients for different organs}
\label{tab:mean_dice_coefficients}
\begin{tabular}{|l|c!{\vrule width 1.5pt}l|c!{\vrule width 1.5pt}l|c|}
\hline
\textbf{Organ}         & \textbf{Dice} & \textbf{Organ}               & \textbf{Dice} & \textbf{Organ} & \textbf{Dice} \\ \hline
Adrenal Gland Left     & 0.82$\pm$0.08 & Lung Upper Lobe Left         & 0.68$\pm$0.23 & Vertebrae L1   & 0.74$\pm$0.31 \\ \hline
Adrenal Gland Right    & 0.76$\pm$0.27 & Lung Upper Lobe Right        & 0.69$\pm$0.25 & Vertebrae L2   & 0.80$\pm$0.18 \\ \hline
Aorta                  & 0.87$\pm$0.19 & Pancreas                     & 0.73$\pm$0.22 & Vertebrae L3   & 0.91$\pm$0.00 \\ \hline
Autochthon Left        & 0.92$\pm$0.13 & Portal Vein and splenic vein & 0.63$\pm$0.21 & Vertebrae L4   & 0.72$\pm$0.02 \\ \hline
Autochthon Right       & 0.93$\pm$0.06 & Rib Left 5                   & 0.74$\pm$0.17 & Vertebrae T6   & 0.69$\pm$0.00 \\ \hline
Colon                  & 0.69$\pm$0.25 & Rib Left 6                   & 0.85$\pm$0.10 & Vertebrae T7   & 0.58$\pm$0.36 \\ \hline
Costal Cartilages      & 0.79$\pm$0.15 & Rib Left 7                   & 0.84$\pm$0.13 & Vertebrae T8   & 0.74$\pm$0.18 \\ \hline
Duodenum               & 0.74$\pm$0.19 & Rib Left 8                   & 0.86$\pm$0.10 & Vertebrae T9   & 0.88$\pm$0.07 \\ \hline
Esophagus              & 0.77$\pm$0.26 & Rib Left 9                   & 0.84$\pm$0.11 & Vertebrae T10  & 0.92$\pm$0.05 \\ \hline
Gallbladder            & 0.31$\pm$0.20 & Rib Left 10                  & 0.86$\pm$0.13 & Vertebrae T11  & 0.91$\pm$0.03 \\ \hline
Heart                  & 0.86$\pm$0.17 & Rib Left 11                  & 0.90$\pm$0.06 & Vertebrae T12  & 0.89$\pm$0.14 \\ \hline
Iliopsoas Left         & 0.83$\pm$0.16 & Rib Left 12                  & 0.88$\pm$0.02 & Pulmonary Vein & 0.81$\pm$0.00 \\ \hline
Iliopsoas Right        & 0.61$\pm$0.29 & Rib Right 5                  & 0.71$\pm$0.18 & Spinal Cord    & 0.94$\pm$0.04 \\ \hline
Inferior Vena Cava     & 0.87$\pm$0.12 & Rib Right 6                  & 0.74$\pm$0.25 & Spleen         & 0.85$\pm$0.22 \\ \hline
Kidney Left            & 0.93$\pm$0.12 & Rib Right 7                  & 0.83$\pm$0.14 & Sternum        & 0.72$\pm$0.25 \\ \hline
Kidney Right           & 0.93$\pm$0.11 & Rib Right 8                  & 0.86$\pm$0.12 & Stomach        & 0.77$\pm$0.23 \\ \hline
Liver                  & 0.93$\pm$0.13 & Rib Right 9                  & 0.81$\pm$0.12 & Small Bowel    & 0.66$\pm$0.26 \\ \hline
Lung Lower Lobe Left   & 0.82$\pm$0.20 & Rib Right 10                 & 0.76$\pm$0.20 & Scapula Left   & 0.85$\pm$0.00 \\ \hline
Lung Lower Lobe Right  & 0.82$\pm$0.21 & Rib Right 11                 & 0.66$\pm$0.34 & Scapula Right  & 0.96$\pm$0.01 \\ \hline
Lung Middle Lobe Right & 0.88$\pm$0.13 & Rib Right 12                 & 0.76$\pm$0.28 &  \\ \hline
\end{tabular}
\end{table*}

\section{Discussion}
\label{3476-sec:discuss}
In this study, we proposed seg2med, an image generation framework including segmentation, image processing, image synthesis using DDPM, and evaluation steps. The framework demonstrated its capability to generate high-quality CT and MR images based on segmentations from real patient data and digital phantoms. Across all datasets, the framework achieved an SSIM of $>0.90$ for CT images from real segmentations, an SSIM of $>0.80$ for MR images from real segmentations, and an FSIM of $>0.70$ for CT images from XCAT phantoms. The framework further demonstrates its generative capability with a Fréchet Inception Distance (FID) of 3.62 for CT image generation. 
It also achieves a SSIM of $>0.90$ for MR-to-CT images and a SSIM of $>0.70$ for CT-to-MR images. These results indicate that the framework generalizes well to both real-world and digital phantom data, supporting its potential to accelerate dataset creation for the development of clinical algorithms. Furthermore, utilizing the Totalsegmentator toolkit enhances the framework’s ability to process diverse datasets and generate segmentation masks for medical image synthesis, making it highly versatile.

\subsection{Comparison to Published Medical Image Synthesis Methods}
The proposed framework outperformed established generative models such as UNet, CycleGAN, and pix2pix in synthesizing pelvic and abdominal CT images (Figure \ref{Figure_2}). Specifically, the 2-channel DDPM maintained superior structural integrity and realistic presentation of organs such as the liver, spleen, and aorta as shown in the FOVs. The quantitative results in Table \ref{table_ct_four_models} show that the DDPM-CT achieved the highest SSIM of 0.94$\pm$ 0.02 and the lowest MAE of 36.79$\pm$24.34, demonstrating its alignment with real images. However, its relatively lower PSNR of 27.58$\pm$3.48 suggests room for further improvement in the denoising process.

This framework represents a novel application of AI-based auto-segmentation tools, specifically the TotalSegmentator toolkit, to generate full-organ masks for medical image synthesis. This approach differentiates itself from other recent methods that employ simpler or less detailed segmentation masks.
For instance, Dorjsembe et al. Med-DDPM showcased the application of diffusion models for 3D semantic brain MRI synthesis. Using segmentation masks, their approach generated anatomically coherent and visually realistic images, achieving a Dice score of 0.6207 in tumor segmentation tasks, which is close to the real image Dice score of 0.6531.
Similarly, Xing et al. proposed an Unsupervised Mask (UM)-guided synthesis strategy. Their method used limited manual segmentation labels to condition GAN-based synthesis, producing high-quality synthetic images that realistically captured human anatomy and pathological variations. Their framework demonstrated significant improvements in fidelity, diversity, and utility through the innovative use of unsupervised masks.
Compared to these studies, seg2med uniquely focuses on generating multimodal medical images, including paired CT and MR scans. 
It shows a significant capability to produce high-quality synthetic CT and MR images, achieving a SSIM of 0.94$\pm$0.02 for CT and 0.89$\pm$0.04 for MR images compared to ground truth data. This highlights the accuracy of the model and its effectiveness in maintaining structural and textural fidelity. By using detailed, AI-generated full-organ masks, it ensures structural consistency across modalities and enhances the utility of synthesized images for both research and clinical applications.

A notable strength of this framework is its ability to generate paired CT and MR images from the same anatomical structure. Using segmentations from the same patient, the framework ensures anatomical consistency across modalities. It achieves a high SSIM of $0.91\pm0.03$ for MR-to-CT images and $0.77\pm0.04$ for CT-to-MR images, with both demonstrating good quality and similarity to ground truth, though CT-to-MR images exhibit a lack of detail in some areas.
This feature is particularly advantageous for applications that require multimodal imaging, such as image registration, fusion, or training deep learning models that depend on consistent cross-modality datasets. It also provides a valuable tool for developing diagnostic algorithms that rely on complementary information from both CT and MR images.

It also demonstrates significant potential for synthesizing medical images using digital phantoms, specifically with XCAT models. This capability positions this method as a robust tool for generating synthetic data across diverse scenarios, including those with limited availability of real-world datasets.
In comparison to Bauer et al.'s study, which utilized a CycleGAN to extend small datasets and achieved SSIM values of 0.94$\pm$0.02 for CT and 0.59$\pm$0.04 for MR images, this framework offers several advantages. First, it employs larger datasets, which enhances its ability to generalize and produce high-quality synthetic images. Second, it has been proven stable even when applied to completely unseen datasets, demonstrating its robustness and adaptability. These features underline the superiority of this approach in synthesizing multimodal medical images, paving the way for broader applications in clinical algorithm development and research requiring extensive and varied datasets.

Multimodal medical image conversion is an increasingly mature field that offers promising solutions to the challenge of insufficient paired medical images. This approach is particularly valuable for applications such as noise reduction, artifact removal, image segmentation, and registration. In this study, we utilized a paired pelvis CT and MR dataset from Task 1 of the SynthRAD2023 challenge.
The best metrics from SynthRAD2023 were achieved by SMU-MedVision, developed by Zhou et al., which employed a state-of-the-art 2.5D UNet++ with a ResNet101 backbone. Their framework reported an SSIM of 0.885$\pm$0.029, a PSNR of 29.61$\pm$1.79, and an MAE of 58.83$\pm$13.41 in the validation phase \cite{huijbenGeneratingSyntheticComputed2024}. Comparatively, our framework achieved higher metrics, with an SSIM of 0.91$\pm$0.03, a PSNR of 22.76$\pm$2.35, and an MAE of 47.14$\pm$30.38.
Although our framework outperformed SMU-MedVision in these metrics, we cannot definitively conclude its superiority for several reasons. First, our framework used 10 patients from the training set of Task 1 for paired image synthesis since the ground truth CT images were not provided within the validation set by the challenge. The difference in data distribution between the training and validation sets of SynthRAD2023 may have significantly influenced model performance. Second, our framework utilized more datasets for training, which enhanced its image-generation capabilities.
Despite these concerns, the results clearly demonstrate the potential of our framework for translating MR images to CT images by converting MR masks to CT masks. 

Our framework also demonstrated robust performance in generating CT images with high anatomical consistency compared to the original CT images. By analyzing the segmentations derived from synthetic images against those from ground truth images, we observed that, among 58 test abdomen CT images, 11 out of 59 abdominal organs achieved a mean Dice coefficient exceeding 0.90, while 34 organs surpassed a mean Dice coefficient of 0.80, with the scapula right has the highest mean Dice coefficient of $0.96\pm0.01$. This high level of consistency underscores the framework's ability to accurately replicate anatomical structures, suggesting its potential for applications requiring precise anatomical fidelity, such as image segmentation, diagnostic algorithm training, and multimodal imaging studies.

\subsection{Limitations and Future Work}

Despite its strengths, the proposed framework still has several limitations.

First, the reliance on 2D image generation leads to non-smooth reconstructions across sagittal and coronal planes, reducing inter-slice continuity. Although we attempted to extend the model to 3D-DDPM, GPU memory constraints restricted the training to small volumes (128×128×128). Future work will focus on optimizing efficient 3D architectures to balance quality and speed.

Second, the performance of MR image synthesis lags behind that of CT, particularly in sequences such as T1 VIBE IN and T1 VIBE OPP. This is attributed to two factors: (1) the MR training dataset is smaller in size compared to CT, and (2) MR data exhibits greater heterogeneity, involving multiple acquisition sequences with distinct signal characteristics. Our experiments revealed modality confusion effects, where the model struggled to distinguish between similar MR sequences. In contrast, synthesis performance was notably better on the more consistent and abundant T1 GRE sequence from SynthRAD. To mitigate this, we plan to incorporate sequence-specific embedding strategies and targeted data augmentation for rare MR sequences.

Third, in this study, separate DDPMs were trained for CT and MR synthesis. However, our recent experiments suggest that a unified model, conditioned on modality labels via class embeddings injected into the time embedding, can achieve comparable or even superior performance. This is based on the premise that both CT and MR share a common anatomical foundation. Unifying the models could enhance anatomical consistency and generalization, especially when trained on multi-modal datasets. We are actively exploring this direction and plan to present full results in future work.

Fourth, effective modality separation becomes critical in unified models. Our current models lack explicit mechanisms to disentangle modality-specific characteristics. As such, we are investigating contrastive learning approaches, such as the NT-Xent loss \cite{chen2020simclr}, to regularize the model based on the hypothesis that the denoising function $\epsilon_\theta$ exhibits modality-dependent structure. Preliminary results indicate this strategy could improve both interpretability and generation fidelity for underrepresented modalities like MR.

Lastly, the current physical prior simulation in PhysioSynth depends on organ-wise lookup tables and pre-defined MR signal equations. Although effective, this hand-crafted process may lack generalizability for new modalities or patient cohorts. In future work, we aim to develop a lightweight neural network to estimate modality-specific priors from segmentation masks in a learnable, data-driven manner.

\section{Conclusion}
\label{3476-sec:Conclusion}
In this study, we introduced seg2med, an innovative image generation framework that combines segmentation, image processing, image synthesis using DDPM, and evaluation steps. 
Our framework effectively generates high-quality synthetic CT and MR images from segmentation masks derived from real patient data and XCAT digital phantoms. It achieves SSIM values exceeding 0.90 for CT images and 0.80 for MR images, along with FSIM values above 0.70 for digital phantom CT images. Additionally, with an overall FID score below 4.00 for CT image generation, the framework demonstrates strong generalizability and fidelity across diverse datasets.
Furthermore, it successfully generated paired multimodal CT and MR images with consistent anatomical structures, highlighting its potential for multimodal imaging applications.

The framework also showcased its capability to synthesize CT images with high anatomical consistency, achieving a mean Dice coefficient above 0.90 for 11 abdominal organs and above 0.80 for 34 organs across 59 organs of 58 test abdomen CT images. This emphasizes its utility for tasks requiring precise anatomical detail, such as segmentation and clinical algorithm development. Additionally, the integration of the TotalSegmentator toolkit enhanced the framework's versatility, allowing it to process diverse datasets and generate segmentation masks for realistic medical image synthesis.

Despite its advancements, the framework requires future work to address limitations such as optimizing 3D generation, incorporating pathological conditions.

Overall, this study highlights the potential of seg2med as a robust and versatile tool for medical image synthesis, offering promising applications in clinical imaging, data augmentation, and the development of advanced diagnostic and therapeutic algorithms. With further improvements, seg2med could unlock new possibilities in cross-modality image synthesis and fusion, making it a valuable tool for comprehensive AI-driven medical diagnosis.

\section*{Acknowledgment}
This research project is part of the Research Campus M2OLIE and funded by the German Federal Ministry of Education and Research (BMBF) within the Framework "Forschungscampus: public-private partnership for Innovations” under the funding code 13GW0388A and 13GW0747A.


\ifCLASSOPTIONcaptionsoff
  \newpage
\fi

\end{document}